%% file: main.tex
\documentclass[letterpaper, 10 pt, conference]{ieeeconf}  % Comment this line out if you need a4paper

\IEEEoverridecommandlockouts                              % This command is only needed if 
\usepackage[utf8]{inputenc}
\usepackage[sorting=none,maxbibnames=99,giveninits]{biblatex} %Imports biblatex package
\usepackage{graphicx} % Required for inserting images
\usepackage{adjustbox}
\usepackage{amsmath}
\usepackage{amsthm}
\usepackage{amssymb}
\usepackage{tabularx}
\usepackage{empheq}
\usepackage{makecell}
\usepackage{array}

\newtheorem{problem}{Problem}
\newtheorem*{remark}{Remark}

\usepackage{mathtools}

\newcommand{\defeq}{\vcentcolon=}
\usepackage{multirow}
\usepackage{hhline}
\usepackage{amsfonts} 
\usepackage{arydshln}

\usepackage{accents}
\newcommand{\ubar}[1]{\underaccent{\bar}{#1}}
\usepackage{algorithm}
\usepackage{algpseudocode}

\DeclareMathOperator*{\argmin}{arg\,min}
\newcolumntype{?}{!{\vrule width 1pt}}

\title{\LARGE \bf
Safe Control of Quadruped in Varying Dynamics via\\Safety Index Adaptation
}

\author{Kai S. Yun$^{1}$, Rui Chen$^{2}$, Chase Dunaway$^{3}$, John M. Dolan$^{2}$, Changliu Liu$^{2}$%
\thanks{$^1$Mechanical Engineering Department, Carnegie Mellon University, Pittsburgh, PA. Contact: \texttt{sirkhooy@andrew.cmu.edu}}%
\thanks{$^2$Robotics Institute, Carnegie Mellon University, Pittsburgh, PA. Contact: \texttt{\{rui3, jdolan, cliu6\}@andrew.cmu.edu}}%
\thanks{$^3$Mechanical Engineering Department, New Mexico Institute of Mining and Technology, Socorro, NM. Contact: \texttt{chase.dunaway@student.nmt.edu}}%
}

\begin{document}

\maketitle
\thispagestyle{empty}
\pagestyle{empty}

%%%%%%%%%%%%%%%%%%%%%%%%%%%%%%%%%%%%%%%%%%%%%%%%%%%%%%%%%%%%%%%%%%%%%%%%%%%%%%%%
\begin{abstract}
Varying dynamics pose a fundamental difficulty when deploying safe control laws in the real world. Safety Index Synthesis (SIS) deeply relies on the system dynamics and once the dynamics change, the previously synthesized safety index becomes invalid. In this work, we show the real-time efficacy of Safety Index Adaptation (SIA) in varying dynamics. SIA enables real-time adaptation to the changing dynamics so that the adapted safe control law can still guarantee 1) forward invariance within a safe region and 2) finite time convergence to that safe region. This work employs SIA on a package-carrying quadruped robot, where the payload weight changes in real-time. SIA updates the safety index when the dynamics change, e.g., a change in payload weight, so that the quadruped can avoid obstacles while achieving its performance objectives. Numerical study provides theoretical guarantees for SIA and a series of hardware experiments demonstrate the effectiveness of SIA in real-world deployment in avoiding obstacles under varying dynamics.
\end{abstract}

%%%%%%%%%%%%%%%%%%%%%%%%%%%%%%%%%%%%%%%%%%%%%%%%%%%%%%%%%%%%%%%%%%%%%%%%%%%%%%%%

% \section{INTRODUCTION}
\section{Introduction}
\input{sections/introduction}

%%%%%%%%%%%%%%%%%%%%%%%%%%%%%%%%%%%%%%%%%%%%%%%%%%%%%%%%%%%%%%%%%%%%%%%%%%%%%%%%
% \section{RELATED WORK}
\section{Related Work}
\input{sections/related_work}

%%%%%%%%%%%%%%%%%%%%%%%%%%%%%%%%%%%%%%%%%%%%%%%%%%%%%%%%%%%%%%%%%%%%%%%%%%%%%%%%
% \section{PRELIMINARIES}
\section{Preliminaries}
% In this section, we define the problem of safe control under varying dynamics. We further outline system assumptions and our definition of safety. Then we introduce the SIS and SIA problems.
% In this section, we define the problem of safe control under varying dynamics and introduce the SIS and SIA problems.
\subsection{Safe Control for Varying Dynamics}
\input{sections/prelim_dynamic_system}

\subsection{Safety Index Synthesis}
\input{sections/prelim_sis}

\subsection{Safety Index Adaptation via DGA}
\input{sections/prelim_sia}

%%%%%%%%%%%%%%%%%%%%%%%%%%%%%%%%%%%%%%%%%%%%%%%%%%%%%%%%%%%%%%%%%%%%%%%%%%%%%%%%
% \section{PROBLEM FORMULATION}
\section{Problem Formulation}
The system dynamics of the quadruped change with the package weight it carries. We first synthesize an initial safety index for the quadruped dynamics without payload. Then, SIA adapts this safety index for varying payloads. 
To effectively abstract out the complex dynamics and apply the adaptive safe control, we introduce the extended unicycle model and derive the SIS and SIA for the system in this section.
% This section defines the varying dynamics based on the extended unicycle model and derives SIS and SIA for the system.
% The quadruped has a second-order dynamics model with body accelerations and yaw rate as its commands. 

\subsection{Extended Unicycle Dynamics}\label{sec:ext_unicycle_dynamics}
\input{sections/prob_ext_unicycle}

% \subsection{Affine Parameter-varying Dynamic System}
% \input{sections/prob_affine_parameter}

\subsection{SIS and SIA for Varying Quadruped Dynamics}
\input{sections/prob_sis_quadruped}

%%%%%%%%%%%%%%%%%%%%%%%%%%%%%%%%%%%%%%%%%%%%%%%%%%%%%%%%%%%%%%%%%%%%%%%%%%%%%%%%
% \section{HARDWARE SETUP AND SYSTEM IDENTIFICATION}
\section{Hardware Setup and System Identification}

Naturally, the first step in safety index adaptation is identifying the varying parameters for different payloads, enabling DGA.
% , enabling determinant gradient ascent. 
We use payloads of 0.0kg (empty basket), 3.5kg, and 5.9kg. 
% , and identify the varying parameters for each. 
This section details the hardware, system identification process, and the identified varying parameters.

% \footnote{To carry the packages, a basket is attached to the top of the quadruped as shown in Fig.~\ref{fig:extended_unicycle_dyn}. Thus, we identify the system under no payload.}

\subsection{Unitree Go2 Quadruped}
\input{sections/sysid_unitreeGo2}

\subsection{System Identification for Varying Parameters}\label{sec:sysid}

\input{sections/sysid_varying_parameters}
\section{Experiments}

\begin{figure*}[htbp]
    \centering
    \begin{minipage}[t]{0.53\textwidth}
        \vspace*{+0.4\baselineskip}
        \includegraphics[width=\textwidth]{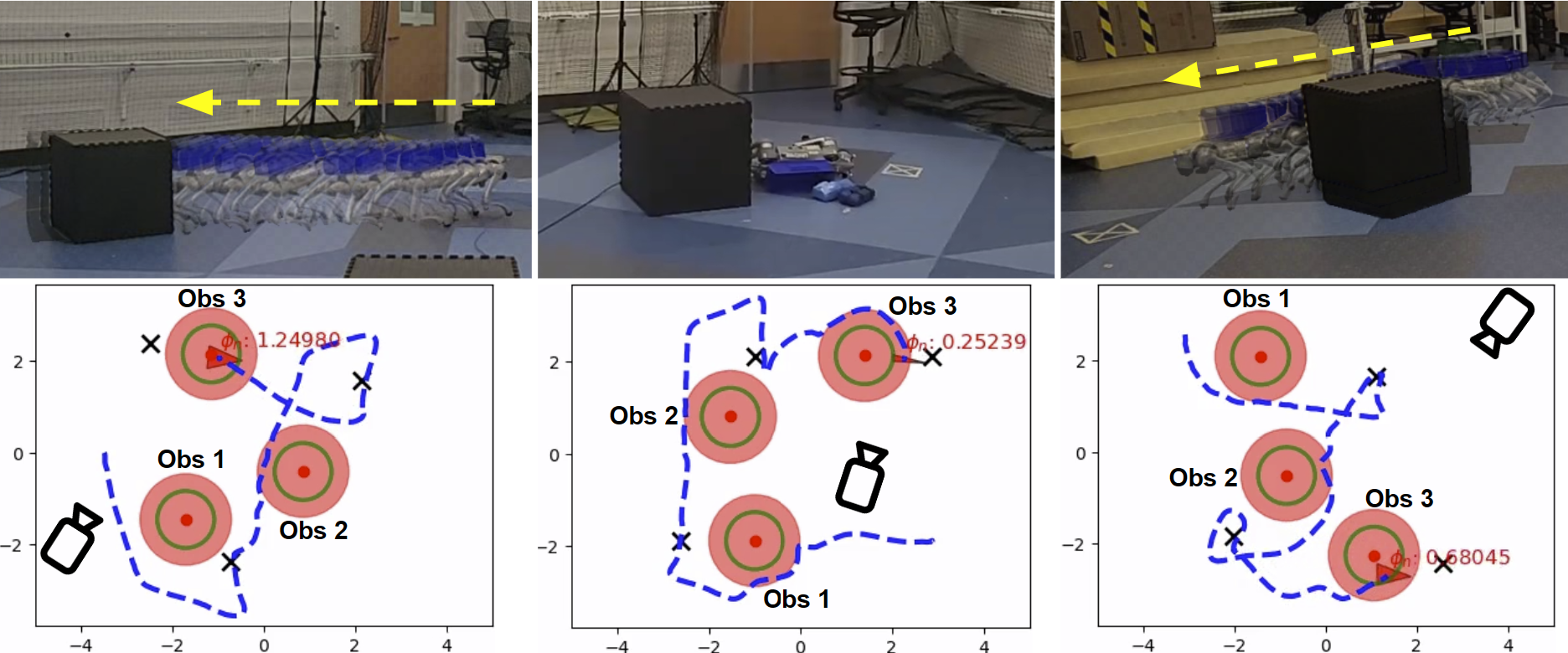}
    \end{minipage}%
    \hspace{0.2cm}
    \begin{minipage}[t]{0.35\textwidth}
        % \vspace*{\baselineskip}
        \caption{Failure cases using a non-adapted safety index for heavy payloads. From left to right: Course 1, 2, 3. The top row shows motion trail images of the failure cases. The bottom row shows the trajectories at failures using motion capture data. In the bottom row, the triangle represents the quadruped, the dotted blue line is its trajectory, the dark red dot marks the obstacle, the light red circle is the user-defined unsafe set $\mathcal{X}_S^C$, and the green solid line is the unsafe set boundary padded with discrete-time safety margin, $\sigma_{\text{DT}}$. The $\phi_\theta$ value is displayed above the quadruped.}
    \end{minipage}
    % \vspace{-0.5cm}
\end{figure*}

We first conduct a numerical study on safety indices produced by SIS and SIA for varying parameters across different packages. 
% Here, the feasibilities of the synthesized and adapted safety indices are validated by sample-based evaluations. 
We then deploy the safety indices on a quadruped to validate their real-time performance. The quadruped, carrying different payloads, adapts its safety index and maneuvers around obstacles to reach target locations. % while maintaining balance.
Both numerical and hardware experiments reveal failure cases for the non-adapted safety index. Using SIA, the safety objectives are consistently met.

\subsection{Numerical Study}
\input{sections/numstud_feas_analysis}

\subsection{Package-carrying Quadruped Hardware Experiments}\label{sec:hardware_experiments}
\input{sections/hardexp_results}

\subsection{Discussion}\label{sec:discussion}
\input{sections/results_discussion}

%%%%%%%%%%%%%%%%%%%%%%%%%%%%%%%%%%%%%%%%%%%%%%%%%%%%%%%%%%%%%%%%%%%%%%%%%%%%%%%%
\section{Conclusion and Future Work}
In this paper, we improve the efficiency of SIA and demonstrate its ability to guarantee safety for quadruped robots in varying dynamics. 
% While non-adapted safety index fails to provide safety guarantees under varying dynamics, the adapted safety indices successfully avoid all obstacles. 
We also analyze the DT safety margin and its impact, demonstrating that SIA guarantees FI with the margin, while non-adapted cases still fail. We plan to synthesize and adapt a DT safety index for deployment in future work. Additionally, we aim to incorporate an online estimator for the varying parameters.
% Additionally, we aim to provide a theoretical analysis of SIA's effectiveness in finding local optima and the limitations of SIS concerning system dimensions.


%%%%%%%%%%%%%%%%%%%%%%%%%%%%%%%%%%%%%%%%%%%%%%%%%%%%%%%%%%%%%%%%%%%%%%%%%%%%%%%%

%%%%%%%%%%%%%%%%%%%%%%%%%%%%%%%%%%%%%%%%%%%%%%%%%%%%%%%%%%%%%%%%%%%%%%%%%%%%%%%%

%%%%%%%%%%%%%%%%%%%%%%%%%%%%%%%%%%%%%%%%%%%%%%%%%%%%%%%%%%%%%%%%%%%%%%%%%%%%%%%%
% \section*{Appendix}
% % Appendixes should appear before the acknowledgment.
% \input{sections/appendix}
\section*{Acknowledgement}

The authors would like to thank Weiye Zhao at Carnegie Mellon University for technical guidance. Chase Dunaway is supported by the National Science Foundation (NSF) under Grant No. 1950811.

%%%%%%%%%%%%%%%%%%%%%%%%%%%%%%%%%%%%%%%%%%%%%%%%%%%%%%%%%%%%%%%%%%%%%%%%%%%%%%%%

\AtNextBibliography{\small}
% \begin{thebibliography}

\printbibliography

\end{document}

%% file: sections/introduction.tex
In this work, we study the safe control of quadruped robots in a realistic scenario: carrying objects of varying weights. Quadrupeds have been studied for real-world tasks such as helping the visually-impaired~\cite{xiao2021GuideDog}, towing objects collaboratively~\cite{yang2022cable}, and even performing parkour stunts~\cite{chen2024parkour}. When deploying robots in real-world settings~\cite{liu2022jerk, lin2017real,he2024agilesafelearningcollisionfree, liu2023roboticlegoassemblydisassembly,shek2023learn}, it is essential to ensure safety.
% , such as collision avoidance.
In particular, safety guarantees under varying dynamics are of paramount interest.
% in deployment. 
For a shopping companion quadruped, the payload on its back will constantly change as the human loads it with different products. 
Under such changing dynamics, how can the quadruped maintain safety, i.e., avoid collisions? This is the exact question that this work addresses.

We achieve this by using safety index adaptation (SIA)~\cite{chen2024sia}. SIA allows safe control laws to adapt to changing dynamics within seconds, mitigating safety risks from dynamic mismatches. SIA is a member of the safe set algorithm (SSA) family~\cite{liu2014control}, a widely studied method to ensure constraint satisfaction for dynamic systems. SSA quantifies safety using energy functions and derives safe control laws that keep the system states within a user-specified safe region, ensuring forward invariance (FI), meaning the system stays in the safe region once entered~\cite{chen2023safetyindexsynthesisstatedependent}.
While this is sufficient for safety guarantees when the system starts in a safe state, practical issues such as discretization errors and hardware noise may cause the system to drift into unsafe regions when the robot operates near the safe region boundary.
To address this, the control law must also ensure finite-time convergence (FTC), bringing the system back to safety~\cite{chen2023safetyindexsynthesisstatedependent}.
% , as proved to be satisfied by SSA~\cite{chen2023safetyindexsynthesisstatedependent}.
Although safety index synthesis (SIS)~\cite{chen2023safetyindexsynthesisstatedependent, zhao2022safety} provides a systematic method of constructing feasible and safe control laws, it does not provide such guarantees when the dynamics change.
% , which often occurs in real life. 
Reapplying SIS for each new dynamics is impractical, as it can take several minutes, even for simple dynamics.
To handle this, safety index adaptation (SIA)~\cite{chen2024sia} has been proposed.

 \begin{figure}[t]
    \centering
    \includegraphics[width=0.48\textwidth]{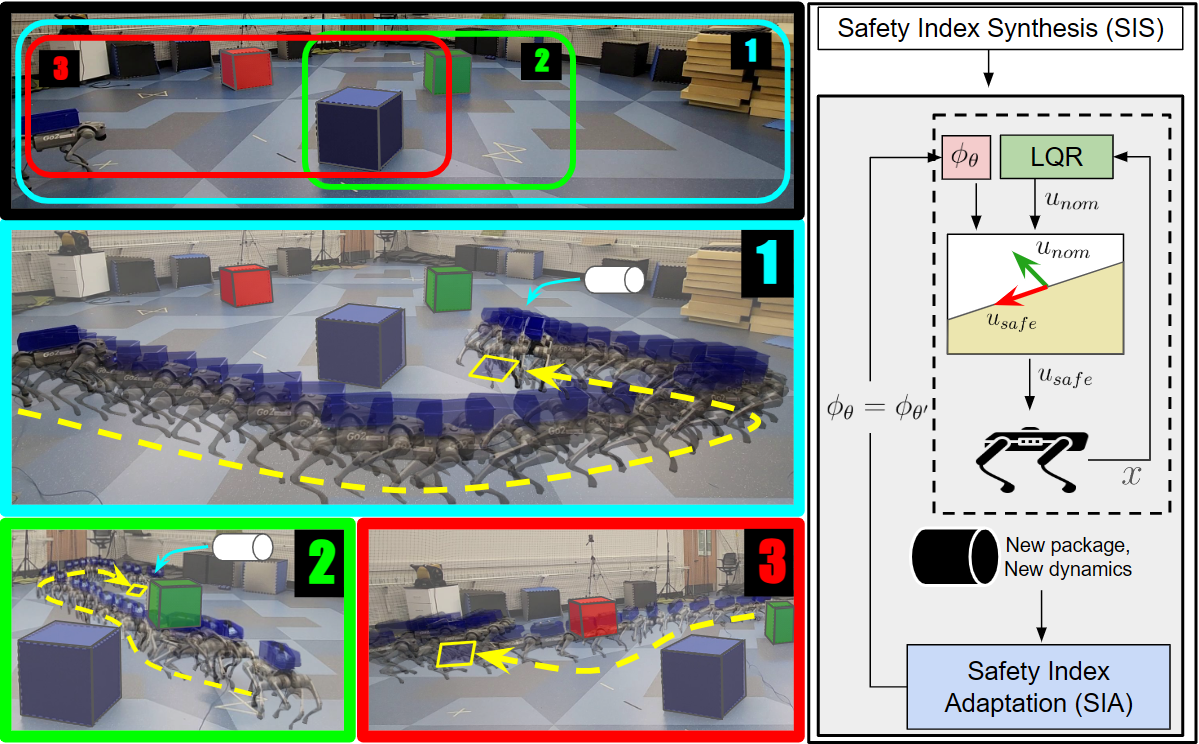}
    \caption{Trajectory of the quadruped carrying various packages of different weights (left column) and overall pipeline (right column). The quadruped avoids each obstacle sequentially, in the order of (1) blue, (2) green, and (3) red cubes, using safety index. Once the quadruped arrives at a goal position after avoiding an obstacle, a different package is loaded on its back. At this point, the safety index is adapted to the new dynamics.}
    \label{fig:full_traj}
    % \vspace{-0.936cm}
\end{figure}

Here, we test SIA on physical robots for the first time and demonstrate the effectiveness of SIA in changing real-world dynamics. 
Specifically, we apply SIA to a quadruped robot tasked with navigating multiple goals while carrying varying payloads and avoiding obstacles. 
We demonstrate that SIA is essential to avoid safety failures under varying dynamics 
% , e.g., avoid collision while maintaining balance, 
% and fast enough for online safe control law adaptation.
and fast enough for online adaptation.
To summarize, our contributions are:
\begin{itemize}
    \item We guarantee safety for real quadruped robots in navigation tasks with varying dynamics.
    \item We improve the efficiency of safety index adaptation and validate its efficacy in real life.
    \item We provide a comprehensive analysis of the safety characteristics of adapted safety indices.
    \item We introduce a linear model for the varying quadruped dynamics and a method to identify varying parameters.
\end{itemize}

%% file: sections/related_work.tex
% Synthesizing a valid safe control law for an arbitrary constrained control problem is non-trivial since it needs to ensure that, at every state, a control input within the limits exists to satisfy the safe control constraints to make the safe control law realizable. 
% To this end, SIS has been developed, using sum-of-squares programming (SOSP) to construct feasible and safe control laws. ~\cite{zhao2022safety} tackles SIS for known systems with control bounds, while~\cite{chen2023safetyindexsynthesisstatedependent} extends this to state-dependent control bounds. 
Formal SIS uses sum-of-squares programming (SOSP) to construct feasible and safe control laws. ~\cite{zhao2022safety} tackles SIS for known systems with control bounds, while~\cite{chen2023safetyindexsynthesisstatedependent} extends this to state-dependent control bounds. 
Another line of work in safe control is control barrier functions (CBF)~\cite{ames2014control, ames2019control}, which also uses energy functions to quantify safety.
While SIS is a subset of CBF synthesis~\cite{clark2022semi, dai2023convex, kang2023verification, liu2022inputsat, dawson2022safecontrollearnedcertificates}, SIS targets a specific class of energy functions usually for collision avoidance and considers both FI and FTC.
% CBF synthesis has also been extensively investigated~\cite{clark2022semi, dai2023convex, kang2023verification, liu2022inputsat, dawson2022safecontrollearnedcertificates}.
% While learning-based synthesis~\cite{ma2022joint,liu2022inputsat,dawson2022safecontrollearnedcertificates,so2023trainneuralcontrolbarrier} have been explored, they lack the mathematical rigor necessary for safety guarantees. 

As previously mentioned, real-world dynamics are imperfectly known and usually change over time.~\cite{xu2015rCBF,jankovic2018robust} propose robust CBFs (rCBFs) to provide safety under bounded uncertainties, while~\cite{ames2020aCBF} develops adaptive CBFs (aCBFs) to estimate the changes in the dynamics. 
% The intersection of these two, robust-adaptive CBFs (raCBFs) has also been studied~\cite{lopez2023unmatched}.
% There are two main paradigms in safe control under such dynamics: \textit{robust} safe control aims to provide safety under bounded uncertainties~\cite{xu2015rCBF,jankovic2018robust} and \textit{adaptive} safe control estimates the changes in the dynamics~\cite{ames2020aCBF}. The intersection of these two, \textit{robust-adaptive} safe control, has also been studied~\cite{lopez2023unmatched}. 
% Adaptive safe control is less conservative compared to robust safe control since they do not consider the ``worst-case" scenarios.
Although safe controller synthesis for constant dynamics has been studied extensively, synthesis of adaptive safe controllers under changing dynamics has not been studied widely to our knowledge.~\cite{sliu2024racbf} uses SOSP to synthesize a robust-adaptive CBF (raCBF) that applies adaptive safe control for systems with bounded parametric uncertainties.~\cite{fan2020bayesian,brunke2022barrierbayesianlinearregression} assume bounded dynamics noise for its aCBF, but lack the mathematical rigor required for safety guarantees since they use learning-based methods.~\cite{chen2024sia} leverages Sylvester's criterion to readily adapt safety index to varying dynamics while guaranteeing safety constraints. It provides the central theory and methodology for our work, but has only been tested in simulation for a 2-DOF robot. For safe control of quadruped hardware,~\cite{he2024agilesafelearningcollisionfree} uses Hamilton-Jacobi-based safeguard for agile locomotion, but does not provide safety guarantees. While ~\cite{grandia2021cbfmpc} uses CBF for rough terrain locomotion and~\cite{liao2023narrow} utilizes CBF to walk through narrow spaces, these do not consider varying dynamics.
This paper addresses the growing interest in providing \textit{adaptive safety assurance for quadrupeds}.

% \textbf{Safe Control of Quadrupeds:} These works involve deploying energy-function-based safe controllers on quadruped platforms in real life. While some work exists in deploying CBFs~\cite{liao2023walkingnarrowspacessafetycritical}, we are unaware of any previous work deploying adaptive safe controllers on quadruped hardware.

%% file: sections/prelim_dynamic_system.tex
The system state is $x\in\mathcal{X}\subseteq \mathbb{R}^{N_x}$ and the control input is $u\in\mathcal{U}\subseteq \mathbb{R}^{N_u}$. The state-space $\mathcal{X}$ is bounded with inequalities, i.e., ${\mathcal{X} = \{ x: h_i(x)\ge 0, \forall i=1,\ldots,N_h \}}$ and the control space $\mathcal{U}$ is bounded element-wise, i.e., $\mathcal{U}=\{ u: \ubar{u} \le u \le \bar{u}\}$. The nominal dynamics, without concern for the changes in dynamics is $\dot x_{nom} = f(x) + g(x) u, \;u\in\mathcal{U}$
% \begin{equation}\label{eq:nominal_dynamics}
%     \dot x_{nom} = f(x) + g(x) u, \;u\in\mathcal{U},
% \end{equation}
with $f:\mathbb{R}^{N_x}\rightarrow\mathbb{R}^{N_x}$ and $g: \mathbb{R}^{N_x}\rightarrow\mathbb{R}^{N_x \times N_u}$, both locally Lipschitz continuous.
% Version 2: short and sweet
In practice, system dynamics are usually varying, and it is critical to parameterize the changing dynamics and incorporate their changes with the safe control law. To this end, we denote the varying parameters as $\rho$ and extend the nominal dynamics, following \cite{chen2024sia}. In this work, we consider the following parameter-varying dynamics:
\begin{equation}\label{eq:affine_varying_dynamics}
    \dot x = f(x,\rho) + g(x,\rho) u = A^f f(x) + A^g g(x) u + \epsilon,
    % \begin{split}
    %     \dot x &= f(x,\rho) + g(x,\rho) u\\
    %     & = A^f f(x) + A^g g(x) u + \epsilon,
    % \end{split}
\end{equation}
where {\small${A^f := \text{diag}([\rho^f_{11},\rho^f_{22}, \ldots, \rho^f_{N_xN_x}]})$} is a real diagonal matrix, {\small${A^g \in\mathbb{R}^{N_x \times N_x}}$} is a free-form real matrix, and {\small${\epsilon = [\rho^{\epsilon}_1, \rho^{\epsilon}_2, \ldots, \rho^{\epsilon}_{N_x}]^T}$} is a real vector. 
% We assume that $\rho$ is known throughout the deployment

% Version 1: Longer and more detailed
% In practice, system dynamics in the real world are usually varying, and it is critical to parameterize the changing dynamics and incorporate their changes with the safe control law. To this end, we denote the varying parameters as $\rho$ and extend (\ref{eq:nominal_dynamics}), following \cite{chen2024sia}:
% \begin{equation}\label{eq:varying_dynamics}
%     \dot x = f(x,\rho) + g(x,\rho) u, \;u\in\mathcal{U}(\rho),
% \end{equation}
% where we assume that $\rho$ is known throughout the deployment.

% In this work, we consider the following affine parameter-varying dynamic system, which extends (\ref{eq:varying_dynamics}) to an affine structure with respect to the varying parameters.
% \begin{equation}\label{eq:affine_varying_dynamics}
%     \dot x = f(x,\rho) + g(x,\rho) u = A^f f(x) + A^g g(x) u + \epsilon,
% \end{equation}
% where {\small${A^f := \text{diag}([\rho^f_{11},\rho^f_{22}, \ldots, \rho^f_{N_xN_x}]}$} is a real diagonal matrix, {\small${A^g \in\mathbb{R}^{N_x \times N_x}}$} is a free-form real matrix, and {\small${\epsilon = [\rho^{\epsilon}_1, \rho^{\epsilon}_2, \ldots, \rho^{\epsilon}_{N_x}]^T}$} is a real vector.
% , parameterized by ${\rho^g_{ij}\;\forall i=1,\ldots,N_x, j=1,\ldots,N_x}$

We enforce safety with two objectives~\cite{chen2023safetyindexsynthesisstatedependent}: (C.1) \textit{forward invariance (FI)} and (C.2) \textit{finite-time convergence (FTC)}. Given a user-specified \textit{safety specification}, defined with a piecewise-smooth function $\phi_0(x) : \mathcal{X}\rightarrow \mathbb{R}$, forward invariance means that if the state $x$ is already within the \textit{forward invariant set}, $\mathcal{X}_S^*$, which is a subset of the \textit{safe set}, $\mathcal{X}_S=\{x:\phi_0(x)\le 0\}$, it should never leave that set. Finite-time convergence means that if the state $x$ is outside the forward invariant set, it should land in that set in finite time.

Safe set algorithm (SSA)~\cite{liu2014control} provides a method for designing a general $n$\textsuperscript{th} ($n\ge0$) order, linearly parameterized \textit{safety index} $\phi_n$ that can handle general relative degrees ($> 1$) between $\phi_0$ and the control input $u$. 
$\phi_n: \mathcal{X}\rightarrow\mathbb{R}$, is a continuous, piecewise-smooth energy function that measures safety and defines forward invariant set $\mathcal{X}_S^*=\{ x: \phi_n(x)\le 0\}$. It is designed as $\phi_n = \left[\prod_{i=1}^n (1+a_i s)\right]\phi_0$, where $s$ is the differentiation operator. This can be expanded to
\begin{align}\label{eq:si_expansion}
    \phi_n := \phi_0 + \textstyle\sum_{i=1}^n k_i \phi_0^{(i)}
\end{align}
where {\small$\phi_0^{(i)}$} is the $i$\textsuperscript{th} time derivative of $\phi_0$. SSA uses the following optimization as the safe control law:
\begin{align}\label{eq:safe_control_law}
    \mathop{\boldsymbol\min}_{u\in\mathcal U}\mathcal{J}(u) \;\;\mathbf{ s.t. }\;\; \dot\phi_\theta(x,u) < -\eta \text{ if } \phi_\theta(x)\ge0 
\end{align}
where the objective $\mathcal{J}$ is arbitrary. \cite{chen2024sia,chen2023safetyindexsynthesisstatedependent} show that if the following conditions are satisfied, both FI and FTC are guaranteed: (i) the characteristic equation $\prod_{i=1}^n (1+a_i s)=0$ only has negative real roots, (ii) the relative degree from $\phi_0^{(n)}$ to control input is one, and (iii) there always exists a control $u$ that satisfies (\ref{eq:safe_control_law}).
The conditions (i) and (ii) above are easy to hold by design, while (iii) is non-trivial to satisfy.
In the next subsection, we summarize how (iii) is achieved via safety index synthesis.

%% file: sections/prelim_sis.tex
Designing a safety index that achieves the safety guarantees involves choosing the right parameterization for (\ref{eq:si_expansion}). This process is referred to as \textit{Safety Index Synthesis} (SIS):
\begin{problem}\label{prob:sis}
    \textup{(Safety Index Synthesis)}. Find safety index as $\phi_\theta := \phi_0 + \sum_{i=1}^n k_i \phi_0^{(i)}$ with parameter $\theta\in\Theta:=\{[k_1,k_2,\ldots,k_n]:k_i\in\mathbb{R}, k_i\ge0, \forall i\}$, such that
    \begin{align}\label{eq:sis_problem}
        \forall x \in \mathcal{X} \;\;\mathbf{ s.t. }\;\; \phi_\theta(x)\ge0,\; \mathop{\boldsymbol\min}_{u\in\mathcal{U}} \dot\phi_\theta(x,u) < -\eta.
    \end{align}
\end{problem}
The $n$\textsuperscript{th}-order safety index, $\phi_n$, is parameterized by $\theta$ and is referred to as $\phi_\theta$. For clarity, $\phi_n$ and $\phi_\theta$ are used interchangeably in this work. 
% Since Problem~\ref{prob:sis} depends on the dynamics and has infinitely many constraints, i.e., (\ref{eq:sis_problem}) needs to hold for \textit{any} state $x$ \textit{s.t.} $\phi_\theta(x)\ge0$, we need a mathematically robust and systematic approach to solve this problem. 
Given the problem's complexity, the method introduced by~\cite{zhao2022safety} is employed. This method leverages Positivstellensatz~\cite{parrilo2003sosp} to transform Problem~\ref{prob:sis} into a SOSP problem, which is then further converted into a nonlinear programming (NP) problem. Specifically, a refute set is constructed and proved empty by solving the SOSP, with detailed procedures outlined in~\cite{chen2023safetyindexsynthesisstatedependent}. The SOSP finds real-valued coefficients $q_i\in\mathbb{R}$, $p_i\ge0, \forall i\in\mathbb{Z}_+$, such that
\begin{equation}\label{eq:sosp_polynomial}
    \begin{split}
         p_0 =& -1 - \textstyle\sum_{i=1}^{N_\zeta}q_i\zeta_i - p_1\gamma_1-p_2\gamma_2-\cdots -p_N\gamma_N \\
         &\quad -p_{12}\gamma_1\gamma_2-\cdots - p_{12\ldots N}\gamma_1\ldots\gamma_N \in SOS.
    \end{split}
\end{equation}
where $\zeta_i,\gamma_i$ encompass the equality and inequality constraints in the SOSP. Note that $\zeta_i$ and $\gamma_i$ depend on the varying dynamics, including the varying parameters $\rho$. 
The feasible coefficients for ($\ref{eq:sosp_polynomial}$) need to satisfy the positive semi-definite (PSD) decomposition $p_0 = \boldsymbol{x}^\top Q(\theta,\boldsymbol{p},\rho)\boldsymbol{x}$ where $Q(\theta,\boldsymbol{x},\rho)\succeq0$. Let {\small$\boldsymbol{x}\defeq[1,x[1],\ldots,x[N_x],x[1]x[2],\ldots,x[N_x]^d]^\top$} contain all monomials of $x$ with order no higher than $d$, assuming that $p_0$ has degree $2d$, and {\small$\boldsymbol{p}\defeq[p_1',\ldots,p_{N_\zeta}',p_1,p_2,\ldots,p_{012\ldots N}]^T$} be the auxiliary decision variable. With this, we have the final Nonlinear Program (NP), which provides the solution to Problem~\ref{prob:sis}:
\begin{problem}\label{prob:nlp}
    \textup{(Nonlinear Programming)}. Find $\theta\in\Theta$ and $\boldsymbol{p}$ where $\boldsymbol{p}[j]\in\mathbb{R}$ for $j>0$ and $\boldsymbol{p}[j]\ge0$ for $j>N_\zeta$, such that $Q(\theta,\boldsymbol{p},\rho)\succeq 0$.
\end{problem}
The PSD condition can be verified through eigenvalue decomposition or by applying Sylvester's criterion to the determinants~\cite{horn2012matrix}. We utilize the eigenvalue decomposition in SIS and use Sylvester's criterion for SIA, to be introduced in the following subsection.

%% file: sections/prelim_sia.tex
SIA, a novel approach to solving Problem~\ref{prob:sis} introduced in~\cite{chen2024sia}, can efficiently adapt safety index parameter $\theta$ when the dynamics change, i.e., $\rho$ changes to $\rho'$. Instead of performing the entire synthesis process again, SIA searches for a new point $(\theta',\boldsymbol{p}',\rho')$ in the neighborhood of the previous solution to Problem~\ref{prob:nlp}, $(\theta^*,\boldsymbol{p}^*,\rho)$, under the assumption that the change in $\rho$ is bounded, i.e., $||\rho - \rho'||\le\delta$ for some $\delta>0$. Mathematically, this is formulated as:
\begin{problem}\label{prob:sia}
    \textup{(Safety Index Adaptation)}. Find $\phi_{\theta'}$ that solves Problem~\ref{prob:sis} with system parameter $\rho'$ given an originally feasible safety index $\phi_{\theta^*}$, i.e.,  $Q(\theta^*,\boldsymbol{p}^*,\rho)\succeq0$.
    % , given a solution $\phi_{\theta^*}$ to Problem~\ref{prob:sis} with system parameter $\rho$, i.e., solve
    \begin{align}\label{eq:sia_problem}
        \theta',\boldsymbol{p}' = \mathop{\boldsymbol\argmin}_{\theta,\boldsymbol{p}} \mathcal{J}(\theta,\boldsymbol{p})\;\;\mathbf{ s.t. }\;\; Q(\theta,\boldsymbol{p},\rho')\succeq0.
    \end{align}
\end{problem}

In SIA, the constraint in (\ref{eq:sia_problem}) is satisfied via determinant gradient ascent (DGA).
Sylvester's criterion~\cite{horn2012matrix} states that a Hermitian matrix is PSD if and only if all the principal minors are nonnegative. Then, the constraint in (\ref{eq:sia_problem}) is equivalent to
\begin{align}\label{eq:sia_condition}
    \text{Det}[Q(\theta,\boldsymbol{p},\rho')]_{I,I} > 0,\;\;\forall I\subseteq[1,\ldots,M]
\end{align}
where $M$ represents the size of $Q$ and $[Q]_{I,J}$ denotes the submatrix of $Q$ formed by selecting the rows indexed by $I$ and columns indexed by $J$. Given that the principal minors are explicit functions of $\theta$ and $\boldsymbol{p}$, (\ref{eq:sia_condition}) can be satisfied via gradient ascends on these parameters with step size $\lambda$:
\begin{align}\label{eq:dga}
    [\theta,\boldsymbol{p}] &= [\theta,\boldsymbol{p}] + \lambda\delta \\
    \delta &= \nabla_{[\theta,\boldsymbol{p}]} \text{Det}[Q(\theta,\boldsymbol{p},\rho')]_{I^*,I^*}\big|_{\theta=\theta,\boldsymbol{p}=\boldsymbol{p}}
\end{align}
where $\text{Det}[Q(\theta,\boldsymbol{p},\rho')]_{I^*,I^*}$ is the current lowest principal minor with indices $I^*$. SIA via DGA is significantly more time-efficient than the full synthesis and guarantees FI and FTC for $\mathcal{X}_S^*$ with the updated $\phi_{\theta'}$ \cite{chen2024sia}. 
% Thus, SIA via DGA is an optimal method for deploying a safe control law under varying dynamics.

%% file: sections/prob_ext_unicycle.tex
\begin{figure}[htbp]
    \centering
    \includegraphics[width=0.35\textwidth]{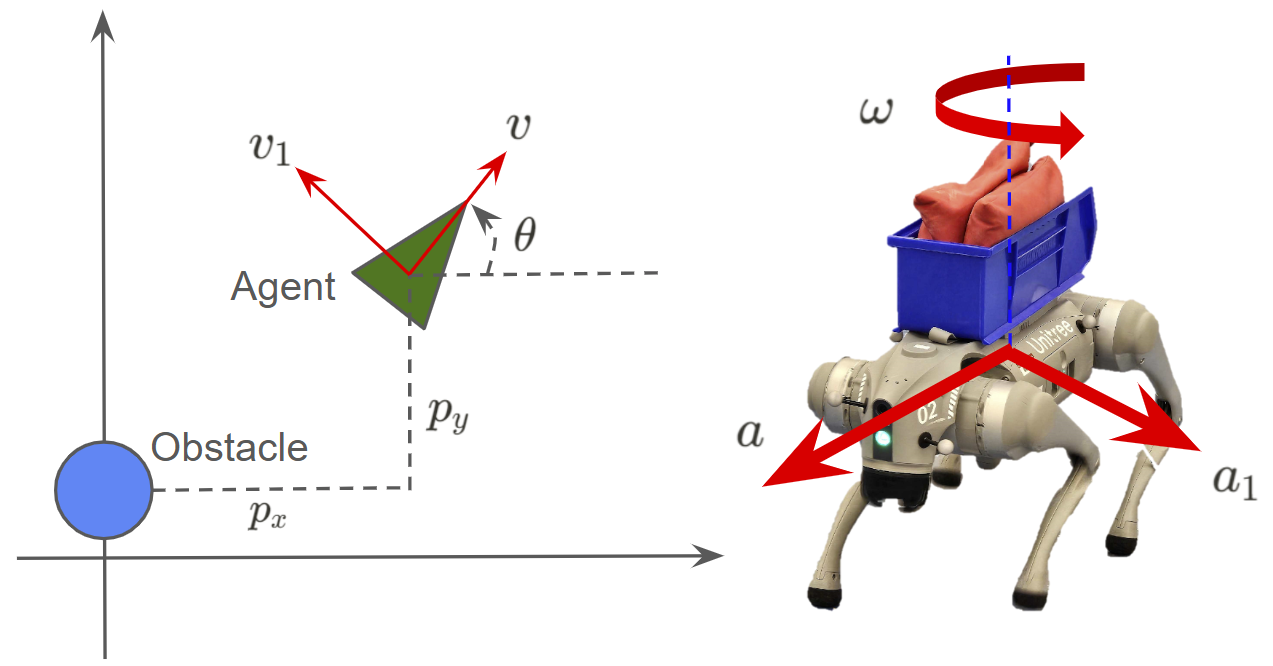}
    \caption{The extended unicycle system models the Go2 quadruped in 2D and allows lateral movements on top of the classical unicycle dynamics.}
    \label{fig:extended_unicycle_dyn}
\end{figure}

For the quadruped dynamics, we follow the parameter-varying dynamic system~(\ref{eq:affine_varying_dynamics}). We introduce the extended unicycle system (Fig.~\ref{fig:extended_unicycle_dyn}), which extends the unicycle system with a quadruped's lateral movements, with states $x=[p_x,p_y,v,v_l,\theta]^\top$ and inputs $u=[a,a_l,\omega]^\top$. 
% {\small
\begin{align}\label{eq:go2_dynamics}
    f(x) = \begin{bmatrix} v\cdot\cos\theta - v_1\cdot\sin\theta \\ v\cdot\sin\theta + v_1\cdot \cos\theta \\ 0 \\ 0 \\ 0 \end{bmatrix},
    g(x) = \begin{bmatrix} 0 & 0 & 0 \\ 0 & 0 & 0 \\ 1 & 0 & 0 \\ 0 & 1 & 0 \\ 0 & 0 & 1  \end{bmatrix}
\end{align}
% }
where $p_x, p_y\in[-1,1] \text{m}$ are the relative positions to the obstacle in the global frame, $v\in[-1.3, 1.3] \text{m/s}$ is the longitudinal body velocity, $v_l\in[-0.7,0.7] \text{m/s}$ is the lateral body velocity, and $\theta$ is the yaw angle in the global frame. $a, a_l\in[-15,15] \text{m/s\textsuperscript{2}}$ are the longitudinal and lateral body acceleration commands and $\omega\in[-2,2]\text{rad/s}$ is the yaw rate command. We set $A^f=\textbf{I}_{5\times 5}$, while $A^g$ and $\epsilon$ take the following form: 
% {\small
\begin{align}\label{eq:ext_var_params}
    A^g = \begin{bmatrix} 1&0&0&0&0 \\ 0&1&0&0&0 \\ 0&0&\rho_{33}^g&\rho_{34}^g&\rho_{35}^g \\ 0&0&\rho_{43}^g&\rho_{44}^g&\rho_{45}^g \\ 0&0&\rho_{53}^g&\rho_{54}^g&\rho_{55}^g \end{bmatrix}, \epsilon = \begin{bmatrix}0\\0\\\rho_3^\epsilon\\\rho_4^\epsilon\\\rho_5^\epsilon\end{bmatrix}.
\end{align}
% }
The varying parameters are identified prior to safety index synthesis and adaptation (Sec.\ref{sec:sysid}). (\ref{eq:go2_dynamics}) and (\ref{eq:ext_var_params}) together form the parameter-varying dynamics (\ref{eq:affine_varying_dynamics}).
% , i.e., $\dot x = A^f f(x) + A^g g(x) u + \epsilon$, used for SIS and SIA of the quadruped in obstacle avoidance. 

%% file: sections/prob_sis_quadruped.tex
We first derive the full SIS solution for the varying quadruped dynamics. We assume that the obstacle is at the origin and the safety specification is {\small $\phi_0 = d_{\min}^2 - d^2$} where {\small $d = \sqrt{p_x^2 + p_y^2}, d_{\min} = $} 1m. Then SIS produces a safety index $\phi_\theta = \phi_0 + k\dot\phi_0$ such that the control law keeps the quadruped's center of mass at least $d_{\min}$ distance away from the obstacle:
\begin{align}\label{eq:safety_index}
    \phi_\theta = d_{\min}^2 - d^2 - 2k d \dot d.
\end{align}
Here, we have a single safety index parameter $k\ge0$. Taking the derivative of $\phi_\theta$, we get the main condition {\small $\mathbf{min}_{u\in\mathcal{U}} \dot\phi_\theta(x,u)<-\eta$} as below
% {\small
\begin{equation}\label{eq:main_feas}
    \begin{split}
        &-2 k v^2 - 2 k v_l^2 - 2 p_x \alpha_1 - 2 p_y \alpha_2 \\
        &-2 k a (-\alpha_3 (\rho_{43}^g + \rho_{53}^g v) + \alpha_4 (\rho_{33}^g - \rho_{53}^g v_l)) \\
        &-2 k a_l (-\alpha_3 (\rho_{44}^g + \rho_{54}v) + \alpha_4 (\rho_{34}^g - \rho_{54}^g v_l)) \\
        &-2 k \omega (-\alpha_3 (\rho_{45}^g + \rho_{55}^g v) + \alpha_4 (\rho_{35}^g - \rho_{55}^g v_l)) < - \eta
    \end{split}
\end{equation}
% }
where $\alpha_i, i=1,2,3,4$ encapsulate the trigonometric terms:
$\alpha_1 = \cos\theta (k\rho_3^\epsilon + v - k\rho_5^\epsilon v_l) - \sin\theta (k\rho_4^\epsilon + v_l + k\rho_5^\epsilon v),    \alpha_2 = \cos\theta (k\rho_4^\epsilon + v_l + k\rho_5^\epsilon v) + \sin\theta (k\rho_3^\epsilon + v - k\rho_5^\epsilon v_l),    \alpha_3 = p_x\sin\theta - p_y\cos\theta,    \alpha_4 = p_x\cos\theta + p_y\sin\theta$.

% simplification terms that encapsulate the trigonometric terms.%used to encapsulate the trigonometric terms:
% {\small
% \begin{align}
%     \alpha_1 &= \cos\theta (k\rho_3^\epsilon + v - k\rho_5^\epsilon v_l) - \sin\theta (k\rho_4^\epsilon + v_l + k\rho_5^\epsilon v) \\
%     \alpha_2 &= \cos\theta (k\rho_4^\epsilon + v_l + k\rho_5^\epsilon v) + \sin\theta (k\rho_3^\epsilon + v - k\rho_5^\epsilon v_l) \\
%     \alpha_3 &= p_x\sin\theta - p_y\cos\theta \\
%     \alpha_4 &= p_x\cos\theta + p_y\sin\theta.
% \end{align}
% }

Since $k\ge0$, the minimum of $\dot\phi_\theta$ is achieved at {\small $a = a_{\max}$} if {\small$ -\alpha_3(\rho_{43}^g+\rho_{53}^gv) + \alpha_4 (\rho_{33}^g -\rho_{53}^gv_l) \ge 0$} and $a = a_{\min}$ otherwise. Similarly, maximum and minimum control inputs are applied to $a_l$ and $\omega$ based on the respective conditions.
% Similarly, $a_l = a_{l,\max}$ achieves the minimum when {\small $ -\alpha_3(\rho_{44}^g+\rho_{54}^gv) + \alpha_4 (\rho_{34}^g -\rho_{54}^gv_l)\ge 0$} and {\small$a_l = a_{l,\min}$} otherwise. Finally, we have {\small$\omega = \omega_{\max}$} if {\small $-\alpha_3 (\rho_{45}^g+\rho_{55}^gv) + \alpha_4 (\rho_{35}^g -\rho_{55}^gv_l) \ge 0$} and {\small$\omega = \omega_{\min}$} otherwise. 
Using indicators $\mathbb{I}_{a,a_l,\omega} = \pm 1$, we can denote these conditions as
% {\small
\begin{align}
    \mathbb{I}_a \left( -\alpha_3(\rho_{43}^g+\rho_{53}^gv) + \alpha_4 (\rho_{33}^g -\rho_{53}^gv_l)\right) &\ge 0 \label{eq:indicator_conditions1}\\
    \mathbb{I}_{a_l} \left( -\alpha_3(\rho_{44}^g+\rho_{54}^gv) + \alpha_4 (\rho_{34}^g -\rho_{54}^gv_l) \right) &\ge 0  \label{eq:indicator_conditions2}\\
    \mathbb{I}_\omega \left( -\alpha_3 (\rho_{45}^g+\rho_{55}^gv) + \alpha_4 (\rho_{35}^g -\rho_{55}^gv_l) \right) &\ge 0 \label{eq:indicator_conditions3}.
\end{align}
% }

In the main feasibility condition (\ref{eq:main_feas}), we replace $a$ with $\Tilde{a}$, $a_l$ with $\Tilde{a}_l$, and $\omega$ with $\Tilde{\omega}$. We have $\Tilde{a}=a_{\max}$ if $\mathbb{I}_a = 1$ and $\Tilde{a}=a_{\min}$ if $\mathbb{I}_a=-1$. Using the same rule, we set the values for $\Tilde{a}_l$ and $\Tilde{\omega}$. We then add conditions for $\alpha_i$'s and limits on states:
{\small
\begin{align}
    -\alpha_1^2 + (k\rho_3^\epsilon + v - k\rho_5^\epsilon v_l)^2 + (k\rho_4^\epsilon + v_l + k\rho_5^\epsilon v)^2 &\ge 0 \label{eq:state_conditions1}\\
    -\alpha_2^2 + (k\rho_4^\epsilon + v_l + k\rho_5^\epsilon v)^2 + (k\rho_3^\epsilon + v - k\rho_5^\epsilon v_l)^2 &\ge 0 \label{eq:state_conditions2}\\
    -\alpha_3^2 + p_x^2 + p_y^2 &\ge 0 \label{eq:state_conditions3}\\
    -\alpha_4^2 + p_x^2 + p_y^2 &\ge 0 \label{eq:state_conditions4}\\
    -p_x^2 + L^2 & \ge 0 \label{eq:state_conditions5}\\
    -p_y^2 + L^2 & \ge 0 \label{eq:state_conditions6}\\
    -v^2 + V^2 & \ge 0 \label{eq:state_conditions7}\\
    -v_l^2 + V_l^2 &\ge 0 \label{eq:state_conditions8}\\
    p_x^2 + p_y^2 - d_{\min}^2 &\ge 0 \label{eq:state_conditions9}
\end{align}
}where $L=1\text{m}$, $V = 1.3 \text{m/s}$, and $V_l = 0.7 \text{m/s}$ are identified as reasonable limits for the states during the system identification procedure (Sec.~\ref{sec:sysid}).

We omit the last condition in (\ref{eq:sis_problem}), $\phi_\theta \ge 0$, to allow the safety index to decrease at all levels, i.e., $\phi_\theta\in\mathbb{R}$, with $\eta = 1e-6$. 
With these, we construct a refute set by negating (\ref{eq:main_feas}) and gathering (\ref{eq:indicator_conditions1})$\sim$(\ref{eq:state_conditions9}), and prove that this refute set is empty. This is equivalent to solving Problem~\ref{prob:sis}~\cite{chen2023safetyindexsynthesisstatedependent}. 
There are 8 versions of the refute set, each assigned with different sign values of $\mathbb{I}_{a,a_l,\omega}$. The $i$\textsuperscript{th} assignment ($i\in[8]$) of $(\mathbb{I}_a,\mathbb{I}_{a_l},\mathbb{I}_\omega)$ leads to the following SOSP problem based on (\ref{eq:sosp_polynomial}):
\begin{align}
    p_{i,0} = \boldsymbol{x}^\top Q_i(\theta,\boldsymbol{p}_i, \rho)\boldsymbol{x} = -1 - \textstyle\sum_{n=1}^{13} p_{i,n}\gamma_{i,n}
\end{align}
where $\boldsymbol{x} := [1,\alpha_1,\alpha_2,\alpha_3,\alpha_4,p_x,p_y,v,v_l]^\top$, $\theta:=[k]$, $\boldsymbol{p}_i:=[p_{i,1},\ldots,p_{i,13}]$, $\rho:=[\rho_{33}^g,\rho_{34}^g,\ldots,\rho_{55}^g,\rho_{3}^\epsilon,\rho_{4}^\epsilon,\rho_{5}^\epsilon]$, and $\gamma_{i,n}$, {\small$\;i\in[1:13]$} encodes inequality constraints (\ref{eq:main_feas})$\sim$(\ref{eq:state_conditions9}).

Once the varying parameters change to $\rho'$, the derivatives of the principal minors of $\{ Q_i\}_{i=1,\ldots,8}$ with respect to $[\theta, \boldsymbol{p}_1,\ldots, \boldsymbol{p}_8]$ are computed to apply (\ref{eq:dga}):
\begin{align}\label{eq:sia_update_quadruped}
    \delta_\theta &= \frac{1}{8}\sum_{i=1}^8 \nabla_\theta \text{Det}[Q_i(\theta,\boldsymbol{p}_i,\rho')]_{I_i^*,I_i^*}\big|_{\theta=\theta, \boldsymbol{p}_i=\boldsymbol{p}_i} \\
    \delta_{\boldsymbol{p}_i} &= \nabla_{\boldsymbol{p}_i}\text{Det} [Q_i(\theta,\boldsymbol{p}_i,\rho')]_{I_i^*,I_i^*}\big|_{\theta=\theta, \boldsymbol{p}_i=\boldsymbol{p}_i}.
\end{align}
We apply the following update rules
% {\small
\begin{align}\label{eq:update_rule}
\theta = \theta + \lambda_\theta\delta_\theta, \;\; \boldsymbol{p}_i = \boldsymbol{p}_i + \lambda_{\boldsymbol{p}}\delta_{\boldsymbol{p}_i}
\end{align}
% }
with learning rates $\lambda_{\theta,\boldsymbol{p}} = 1e-3$ and a discount factor of 0.99 until all principal minors of all $Q_i$'s are below a set target value. Once this is achieved, the safety index parameter $\theta$ is adapted to the new parameters $\rho'$. 
\begin{remark}
    Although gradient updates with a pre-computed symbolic expression of $Q$ are efficient for low-dimensional systems~\cite{chen2024sia}, the size of our $Q$ matrix makes this approach inefficient. Therefore, we utilize finite-differentiation in place of symbolic gradient updates. Both methods yield nearly identical solutions, as the search is limited to a local space.
\end{remark}

%% file: sections/sysid_unitreeGo2.tex
% The Unitree Go2 quadruped (Fig.~\ref{fig:extended_unicycle_dyn}) is controlled with high-level velocity commands $[v^r,v_l^r,\omega^r]$ in the body frame. To translate the acceleration commands in the extended unicycle dynamics (\ref{eq:affine_varying_dynamics}) to velocity commands used by the hardware interface, we perform numerical integration of the acceleration commands:
% \begin{equation}\label{eq:time_integrated_velocity}
%     \boldsymbol{v}_{cmd} = \boldsymbol{v}_{meas}+\boldsymbol{a} dt
% \end{equation}
% where $dt$ is the control period and ${\boldsymbol{a} \defeq [a, a_l]^\top}$, ${\boldsymbol{v}_{cmd}\defeq [v^r, v^r_l]^\top}$, and ${\boldsymbol{v}_{meas} \defeq [v, v_l]^\top}$.  Control inputs are given at 30 Hz for system identification (Sec.~\ref{sec:sysid}) and safety index deployment experiments (Sec.~\ref{sec:hardware_experiments}).

The Unitree Go2 quadruped (Fig.~\ref{fig:extended_unicycle_dyn}) is controlled with high-level velocity commands in the body frame. To translate the acceleration commands in the extended unicycle dynamics (\ref{eq:affine_varying_dynamics}) to velocity commands used by the hardware interface, we perform numerical integration of the acceleration commands:
\begin{equation}\label{eq:time_integrated_velocity}
    \boldsymbol{v}_{cmd} = \boldsymbol{v}_{meas}+\boldsymbol{a}_{cmd} dt,
\end{equation}
where $dt=1/30\text{Hz}$ is the control period. 
% Control inputs are given at 30 Hz for system identification (Sec.~\ref{sec:sysid}) and safety index deployment experiments (Sec.~\ref{sec:hardware_experiments}).
While the Go2 is controlled with high-level commands, the low-level controller is left unaltered. Inevitably, the low-level controller interferes with the high-level commands to rebalance the quadruped. However, when the commands lead to an unsalvagable state, the quadruped falls down.

%% file: sections/sysid_varying_parameters.tex
% We perform a series of system identifications to characterize the quadruped's response to control inputs under different payloads. 

To excite the system, high-level body velocity and yaw rate commands $u_{sid}=[v^{sid},v_l^{sid},\omega^{sid}]$ are generated as sinusoidal waves with various frequencies and amplitudes of $1\text{m/s}, 0.3\text{m/s}, 1.75\text{rad/s}$, for each command. An Optitrack system captures $p_x,p_y,\theta$, while Go2 sensors measure body velocities $v,v_l,\dot \theta$. For data-processing, we convert body velocity commands to desired accelerations by reversing (\ref{eq:time_integrated_velocity}). 

System data is collected for each payload, and smoothed using a LOWESS (Locally Weighted Scatterplot Smoothing) filter. In the data we collected, the quadruped's velocities exceed the specified sinusoidal command amplitudes due to the added weights. Thus, we set state limits for SIS and SIA to be $v\in[-1.3,1.3]\text{m/s}$ and $v_l\in[-0.7,0.7]\text{m/s}$.
Using the filtered data, a least-squares regression is applied to fit $A^g$ and  $\epsilon$ for each package weight. The final varying parameter values are shown in Table~\ref{tab:Ag_values}. 
The identified parameters result in an average coefficient of determination above 0.78. 

\begin{table}[H]
\centering
    \caption{Varying parameters identified from system identification}
    \vspace{-2mm}
    \label{tab:Ag_values}
    {\renewcommand{\arraystretch}{1.0}%
    \begin{adjustbox}{width=0.75\columnwidth}
    % \begin{tabular}{|c||c|c|}
    %      \hline
    %       Payload (kg) & $A^g$ Values ({\small$\rho_{i,j}^g,\forall i,j\in[3:5]$}) & $\epsilon$ Values  \\
    %      \hline
    %       0.0 & $\begin{matrix} 0.08177 &  0.05700 &  -0.00152 \\
    %                             0.00742 &  0.12048 &  0.00241 \\
    %                            -0.00166 &  0.00444 &  0.70741 \end{matrix}$ &
    %      $\begin{matrix} -0.13288 \\ 0.23156 \\  0.01311 \end{matrix}$\\
    %      \hline
    %       3.5 & $\begin{matrix} 0.11144 &  0.02731 &  0.00278 \\
    %                             0.03285 &  0.13207 &  0.00682 \\
    %                            -0.00121 &  0.00207 &  0.68546 \end{matrix}$ &
    %      $\begin{matrix} -0.24451 \\ 0.10565 \\ 0.01923 \end{matrix}$\\
    %      \hline
    %       5.9 & $\begin{matrix} 0.12088 &   0.00613 &   0.00498 \\
    %                             0.04936 &   0.10012 &  -0.03498 \\
    %                             0.00031 &  -0.00129 &   0.66063 \end{matrix}$ &
    %      $\begin{matrix} -0.44301 \\ 0.09005 \\ 0.02785 \end{matrix}$\\
    %      \hline
    %  \end{tabular}
    \begin{tabular}{?c?c|c?}
         \Xhline{2\arrayrulewidth}
          Payload (kg) & $A^g$ Values {\small$\left(\rho_{i,j}^g,\forall i,j\in[3:5]\right)$} & $\epsilon$ Values  \\
         \Xhline{2\arrayrulewidth}
          0.0 & $\begin{matrix} 0.08177 &  0.05700 &  -0.00152 \\
                                0.00742 &  0.12048 &  0.00241 \\
                               -0.00166 &  0.00444 &  0.70741 \end{matrix}$ &
         $\begin{matrix} -0.13288 \\ 0.23156 \\  0.01311 \end{matrix}$\\
         \hline
          3.5 & $\begin{matrix} 0.11144 &  0.02731 &  0.00278 \\
                                0.03285 &  0.13207 &  0.00682 \\
                               -0.00121 &  0.00207 &  0.68546 \end{matrix}$ &
         $\begin{matrix} -0.24451 \\ 0.10565 \\ 0.01923 \end{matrix}$\\
         \hline
          5.9 & $\begin{matrix} 0.12088 &   0.00613 &   0.00498 \\
                                0.04936 &   0.10012 &  -0.03498 \\
                                0.00031 &  -0.00129 &   0.66063 \end{matrix}$ &
         $\begin{matrix} -0.44301 \\ 0.09005 \\ 0.02785 \end{matrix}$\\
         \Xhline{2\arrayrulewidth}
     \end{tabular}
     \end{adjustbox}
     }
\end{table}

%% file: sections/numstud_feas_analysis.tex
We first solve the full SIS for the parameters identified for the 0.0kg payload, $\rho_{0.0}$, in MATLAB using \texttt{fmincon} as the solver. %to get a safety index.
% in MATLAB using \texttt{fmincon} as the solver to get a safety index.
% initialize the quadruped dynamics using the varying parameters identified for the 0.0kg payload, $\rho_{0.0}$, and solve the full SIS (Problem~\ref{prob:nlp}) in MATLAB using \texttt{fmincon} as the solver to get a safety index. 
Then, we further refine the resulting parameters using the update rule (\ref{eq:update_rule}) to achieve the desired safety guarantees for the 0.0kg payload dynamics.
We suspect that the refute set and the number of parameters are too large for the full SIS to converge to an optimal solution before the solver terminates. The update rule (\ref{eq:update_rule}) is effective in converging to a local optimum, refining the safety index to meet the desired safety guarantees. The final initial safety index is $\phi_{0.0}$. 
% Analyses on the limits of full SIS and effectiveness of SIA in finding local optimum are left as future work. 
The update rule (\ref{eq:update_rule}) is then applied for the 3.5kg payload parameters $\rho_{3.5}$ to adapt $\phi_{0.0}$ to $\phi_{3.5}$. Then, $\phi_{3.5}$ is adapted to the 5.9kg payload parameters $\rho_{5.9}$, producing the safety index $\phi_{5.9}$. As seen in Table~\ref{tab:sis_sia_results}, SIA takes significantly less computation time to update the safety index, approximately 1.48\% of the computation time for the full synthesis and refinement combined. 
% Furthermore, we improve the efficiency of SIA by using the central difference method instead of symbolic gradient computation. This results in a 96\% decrease in computation time for SIA.
Furthermore, we improve the computational efficiency of SIA by 96.5\% using finite-differentiation instead of symbolic gradient computation. 

\begin{table}[H]
\centering
    \caption{SIS and SIA Computation Time and Safety Index Parameters}
    \vspace{-2mm}
    \label{tab:sis_sia_results}
    {\renewcommand{\arraystretch}{1.0}%
    \begin{adjustbox}{width=0.75\columnwidth}
    \begin{tabular}{?c?c|c|c?}
         % \hline
         \Xhline{2\arrayrulewidth}
           & $\phi_{0.0}$ & $\phi_{3.5}$ & $\phi_{5.9}$\\
         % \hhline{|=||=|=|=|}
         % \Xhline{2\arrayrulewidth}
         \hline
          $k$ & 0.61068 & 0.64608 & 0.67905\\
         \hline
          Finite-diff. Time (sec) & 562.92 & 9.69 & 7.04\\
         \hline
         % 1923.959926 for SIA after SIS for 0.0kg
          Symbolic-grad. Time (sec) & -- & 398.79 & 76.10\\
         \Xhline{2\arrayrulewidth}
     \end{tabular}
     \end{adjustbox}
     }
\end{table}

Note that the safety index parameter, $k$, increases with heavier payloads (Table~\ref{tab:sis_sia_results}). As discussed in~\cite{chen2024sia}, a system is less sensitive to the inputs when the varying parameters decrease, and thus requires a more aggressive safe control law to react to unsafe regions in advance, i.e., higher $k$ values. In Table~\ref{tab:Ag_values}, observe that {\small$\rho_{5,5}^g$}, 
% the system parameter that maps angular velocity input $\omega$ to {\small$\dot\theta$}, 
decreases as the payload gets heavier. Therefore, the increase in $k$ values for heavier payloads
% , which have lower values for {\small$\rho_{5,5}^g$}, 
is well-aligned with the previous findings.
% Intuitively, the heading angle of the quadruped with respect to the obstacle is a critical indication of safety. Accordingly, the angular velocity input $\omega$ serves as an important input that allows safe control. 

We proceed with sampled-based feasibility experiments.
% quantitative evaluations of the safety indices using sample-based feasibility experiments. 
% The non-adapted safety index, $\phi_{0.0}$, and adapted safety indices, $\phi_{3.5}$ and $\phi_{5.9}$, are used to run the following safe control laws on uniformly sampled states under different payload dynamics:
The non-adapted safety index, $\phi_{0.0}$, and adapted safety indices, $\phi_{3.5}$ and $\phi_{5.9}$, are used to run the following safe control laws under different payload dynamics:
% {\small
\begin{align}
    &\boldsymbol{\min}_{u\in\mathcal U}\mathcal{J}(u) \;\;\mathbf{ s.t. }\;\; \dot\phi_\theta(x,u) \le 0 \text{ if } \phi_\theta(x)<0 \label{eq:feas_test_fi}\\
    &\boldsymbol{\min}_{u\in\mathcal U}\mathcal{J}(u) \;\;\mathbf{ s.t. }\;\; \dot\phi_\theta(x,u) < -\eta \text{ if } \phi_\theta(x)\ge0 \label{eq:feas_test_FTC}
\end{align}
% }

If (\ref{eq:feas_test_fi}) is feasible, the safety index is FI feasible for that state. Similarly, if (\ref{eq:feas_test_FTC}) is feasible, the safety index is FTC feasible. 
% If (\ref{eq:feas_test_fi}) and (\ref{eq:feas_test_FTC}) are feasible, the safety index FI and FTC, respectively, are feasible for that state. 
For each controller-dynamics pair (Table~\ref{tab:num_study_results}), FI and FTC feasibilities are each evaluated for 1000 uniformly sampled states. All the safety indices achieve 100\% feasibility for both FI and FTC in their matching dynamics, e.g., $\phi_{5.9}$ in 5.9kg dynamics. However, $\phi_{0.0}$ fails the FTC condition while the adapted safety indices maintain 100\% feasibility in higher velocity states for 3.5kg and 5.9kg payloads dynamics.
% , highlighting the effectiveness of SIA under varying dynamics.
% {\small
\begin{table}[H]
\centering
    \caption{Non-adapted vs. Adapted Safety Index Feasibility}
    \vspace{-2mm}
    \label{tab:num_study_results}
    {\renewcommand{\arraystretch}{1.0}%
    \begin{adjustbox}{width=\columnwidth}
    % \begin{tabular}{|c||c|c|c||c|c|c|}
    %      \hline
    %        & \multicolumn{3}{c||}{Non-adpated Safety Index} & \multicolumn{3}{c|}{Adapted Safety Index}  \\
    %      \hhline{|=||=|=|=||=|=|=|}
    %       Dynamics & 0.0kg & 3.5kg & 5.9kg & 3.5kg & 5.9kg & 5.9kg\\
    %      \hline
    %       Controller & SI 0.0 & SI 0.0 & SI 0.0 & SIA 3.5 & SIA 3.5 & SIA 5.9\\
    %      \hhline{|=||=|=|=||=|=|=|}
    %      FI & 100\% & 100\% & 100\% & 100\% & 100\% & 100\%\\
    %      \hline
    %      FTC & 100\% & 99.8\% & 99.8\% & 100\% & 100\% & 100\%\\
    %      \hline
    %  \end{tabular}
    \begin{tabular}{?c?c|c|c?c|c|c?}
         \Xhline{2\arrayrulewidth}
           & \multicolumn{3}{c?}{Non-adpated Safety Index} & \multicolumn{3}{c?}{Adapted Safety Index}  \\
         \Xhline{2\arrayrulewidth}
          Dynamics & $\rho_{0.0}$ & $\rho_{3.5}$ & $\rho_{5.9}$ & $\rho_{3.5}$ & $\rho_{5.9}$ & $\rho_{5.9}$\\
         % \hline
         \hdashline 
          Safety Index & $\phi_{0.0}$ & $\phi_{0.0}$ & $\phi_{0.0}$ & $\phi_{3.5}$ & $\phi_{3.5}$ & $\phi_{5.9}$\\
          % Controller & $\phi_{0.0}$ & $\phi_{0.0}$ & $\phi_{0.0}$ & $\phi_{3.5}$ & $\phi_{3.5}$ & $\phi_{5.9}$\\
         \Xhline{2\arrayrulewidth}
         FI (1000 states) & 100\% & 100\% & 100\% & 100\% & 100\% & 100\%\\
         \hline
         FTC (1000 states)& 100\% & 99.8\% & 99.8\% & 100\% & 100\% & 100\%\\
         \Xhline{2\arrayrulewidth}
     \end{tabular}
     \end{adjustbox}
     }
\end{table}
% }

%% file: sections/hardexp_results.tex
% Finally, we validate the effectiveness of SIA in a real-world scenario where the quadruped clears multiple obstacles while carrying different packages. As shown in Fig.~\ref{fig:full_traj}, the quadruped clears each obstacle sequentially and arrives at its desired goal locations. The quadruped starts without payload (0.0kg), and sequentially gets 3.5kg and 5.9kg packages at each goal. At each goal, it waits 5 seconds for package loading, synchronized with SIA computation time for realism. For the baseline, we use the non-adapted safety index $\phi_{0.0}$ throughout the trajectory, while our method changes the safety index, starting with $\phi_{0.0}$ and adjusting to $\phi_{3.5}$ and $\phi_{5.9}$ as dynamics change. Note that SIA is done offline and we implement a switch mechanism to update the safety index with pre-computed values.

We validate SIA's effectiveness in a real-world scenario where the quadruped navigates through obstacles while carrying varying payloads. As shown in Fig.~\ref{fig:full_traj}, the quadruped clears each obstacle and reaches the goal locations sequentially, starting with no payload (0.0kg) and loaded with 3.5kg and 5.9kg packages at the subsequent goals. The safety index considers only one obstacle at a time. The quadruped waits 5 seconds at each goal for package loading, synchronized with SIA computation time for realism. We use the non-adapted safety index $\phi_{0.0}$ as a baseline, while our method adapts the safety index, starting with $\phi_{0.0}$ and adjusting to $\phi_{3.5}$ and $\phi_{5.9}$ as dynamics change. Note that SIA is done offline and we implement a switch mechanism to update the safety index.% with pre-computed values.

% it starts without a package, avoids the first obstacle (blue cube), and reaches the first goal. A 3.5kg package is then loaded, the obstacle changes to the second (red cube), and the quadruped proceeds to the second goal. After reaching the second goal, a 5.9kg package is loaded, and the obstacle switches to the third (green cube). The quadruped then moves to the final goal, completing the experiment. At each goal, it waits 5 seconds for package loading, synchronized with SIA computation time for realism. For the baseline, we use the non-adapted safety index $\phi_{0.0}$ throughout the trajectory, while our method changes the safety index, starting with $\phi_{0.0}$ and adjusting to $\phi_{3.5}$ and $\phi_{5.9}$ as dynamics change. Note that SIA is done offline and we implement a switch mechanism to update the safety index with pre-computed values.

We use three distinct goal-obstacle courses, each repeated multiple times for consistency. 
The varying goal and obstacle positions, shown in the bottom row of Fig. 3, require the quadruped to avoid obstacles from different directions. For instance, in Course 1, it moves forward to the second obstacle and backward to the third, while in Course 3, it moves backward to the second and forward to the third. This variation allows us to evaluate the effectiveness of SIA in real-world scenarios thoroughly. 
LQR controller is used as the nominal controller and the safe control $u$ is computed as follows, 
% through a quadratic projection of the nominal control $u_{\text{nom}}$ when $\phi_\theta\le0$
converting (\ref{eq:safe_control_law}) into a quadratic program (QP)~\cite{ames2014control}: when {\small$\phi_\theta\ge0$,
$u = \argmin_{u\in\mathcal{U}} ||u-u_{\text{nom}}||_2^2\;\;\text{s.t.}\;\;\dot\phi_\theta < -\eta$}.

\begin{table}[H]
\centering
    \caption{Quadruped Hardware Experiment Result}
    \vspace{-3mm}
    \label{tab:hardware_results}
    {\renewcommand{\arraystretch}{1.0}%
    \begin{adjustbox}{width=0.9\columnwidth}
    % \begin{tabular}{|c||c||c|}
    %      \hline
    %      Obstacle Course &  \begin{tabular}{@{}c@{}}Non-adapted Safety Index\\Success \% (Success/Trial)\end{tabular} & \begin{tabular}{@{}c@{}}Adapted Safety Index\\Success \% (Success/Trial)\end{tabular}\\
    %      \hhline{|=||=|=|}
    %      Course 1 & 80\% (4/5) & 100\% (10/10)\\
    %      \hline
    %      Course 2 & 60\% (3/5) & 100\% (6/6)\\
    %      \hline
    %      Course 3 & 80\% (4/5) & 100\% (7/7)\\
    %      \hline
    %  \end{tabular}
    \begin{tabular}{?c?c|c?}
         \Xhline{2\arrayrulewidth}
         Obstacle Course &  \begin{tabular}{@{}c@{}}Non-adapted Safety Index\\Success \% (Success/Trial)\end{tabular} & \begin{tabular}{@{}c@{}}Adapted Safety Index\\Success \% (Success/Trial)\end{tabular}\\
         \Xhline{2\arrayrulewidth}
         Course 1 & 80\% (4/5) & 100\% (10/10)\\
         \hline
         Course 2 & 60\% (3/5) & 100\% (6/6)\\
         \hline
         Course 3 & 80\% (4/5) & 100\% (7/7)\\
         \Xhline{2\arrayrulewidth}
     \end{tabular}
     \end{adjustbox}
     }
\end{table}
% The results from the hardware experiments are presented in Table~\ref{tab:hardware_results}. While the adapted safety indices ensure a 100\% safety rate, meaning the quadruped avoids all obstacles without losing balance, the non-adapted safety index leads to failure cases across all three courses (Fig. 3). In Course 1 (left column), the QP fails when the quadruped reaches its velocity limit ($v=-1.14\text{m/s}$). At this point, the quadruped is moving directly toward the obstacle in reverse, resulting in a collision. In Course 2 (middle column), the quadruped loses balance and falls down while trying to follow a very aggressive safe control input when it's already inside the unsafe region. In Course 3 (right column), the QP remains satisfied, but the non-adapted safety index fails to generate a strong enough control input to meet FTC promptly, leading to another collision. All failure cases happen while carrying the 5.9kg package. 
Table~\ref{tab:hardware_results} shows that the adapted safety indices achieve 100\% safety, ensuring the quadruped avoids all obstacles without losing balance. In contrast, the non-adapted safety index fails to satisfy the safety guarantees across all three courses when carrying the 5.9kg package (Fig. 3). In Course 1, the QP fails when the quadruped nears its velocity limit ($v=-1.14\text{m/s}$), causing a collision as it moves directly toward the obstacle in reverse. In Course 2, the quadruped loses balance and falls while trying to track an overly aggressive control input in an unsafe region. In Course 3, the non-adapted safety index does not provide sufficient control input to ensure prompt FTC, leading to another collision. %All failures occur when carrying the 5.9kg package.

%% file: sections/results_discussion.tex
\begin{figure}[!htbp]
    \centering
    \includegraphics[width=0.47\textwidth]{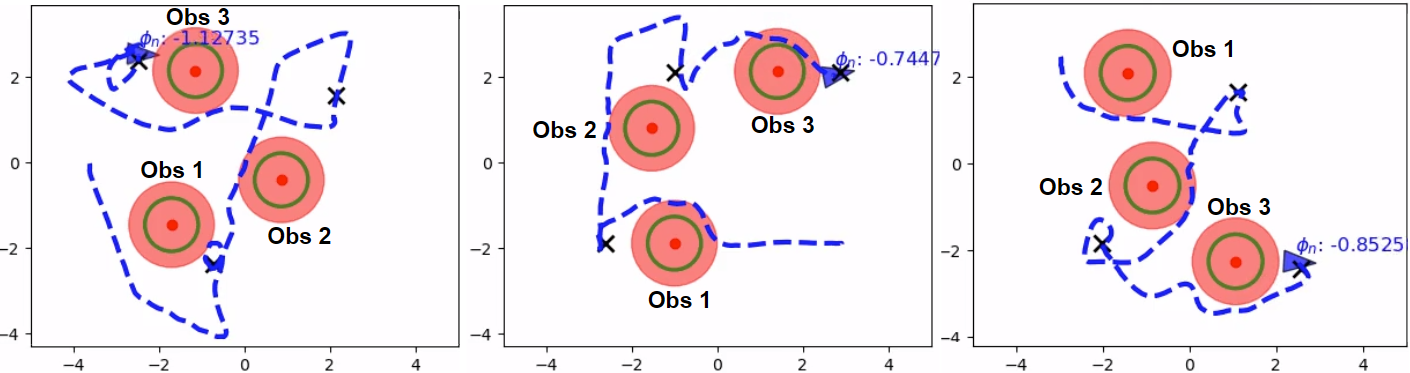}
    \vspace{-2mm}
    \caption{Full trajectories using SIA in Courses 1, 2, and 3.}
    \label{fig:sia_success}
\end{figure}

In this work, we deploy a continuous-time (CT) safe control law in a discrete-time (DT) real system. SSA offers a modified DT safe control law that accounts for discretization errors~\cite{zhao2024implicitsafesetalgorithm}, by incorporating a safety margin $\sigma\in\mathbb{R}_+$ in the safety index formulation: $\phi_\theta = \sigma + d_{\min}^2 -d^2 -2kd\dot d$. Without this margin, the adapted safety indices enter the unsafe set $\mathcal{X}_S^C$, illustrated as light red circles in Fig.~\ref{fig:sia_success}, violating the FI guarantee in practice. To examine whether the deployed safety indices would violate the FI guarantee with a DT safety margin, we derive the practical margin from our hardware experiments: $\sigma_{\text{DT}} = \Delta d_{\max}^2 + 2\Delta d_{\max} \cdot d_{\min}$, where $\Delta d_{\max}=0.375\text{m}$ is the largest change in quadruped position over one timestep. 
Adding $\sigma_{\text{DT}}$ indicates that the radius of the unsafe set is padded with $\Delta d_{\max}$.
We account for the discretization error by padding the safety index with this value (shown with green solid lines in Fig. 3 and Fig.~\ref{fig:sia_success}). With $\sigma_{\text{DT}}$, we see that the adapted safety indices never enter the padded unsafe sets (Fig.~\ref{fig:sia_success}), indicating that the FI is guaranteed, while the non-adapted safety index still violates the FI guarantee (Fig. 3).

\begin{figure}[!htbp]
    \centering
    \includegraphics[width=0.40\textwidth]{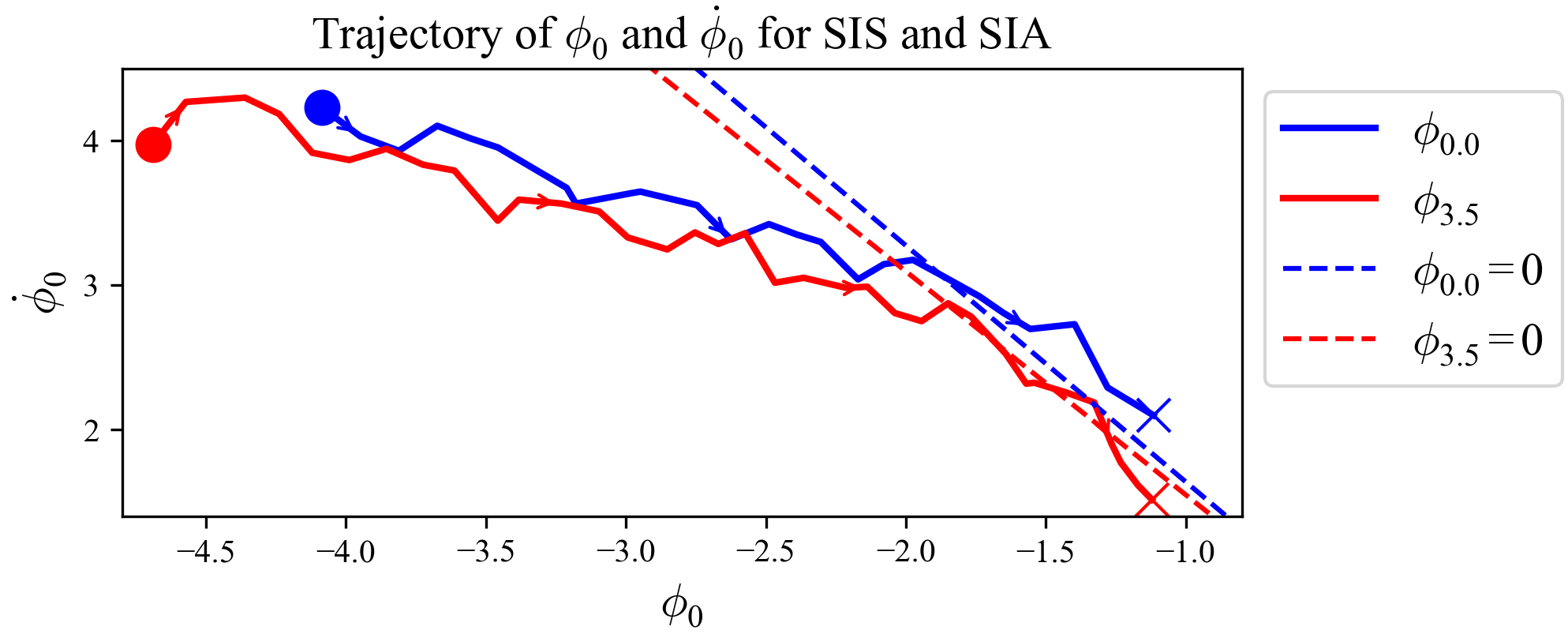}
    \vspace{-2mm}
    \caption{Trajectories (solid lines) of $\phi_{0}$ and $\dot\phi_{0}$ for the non-adapted and adapted safety indices from hardware deployment. The segments move from solid circles to crossmarks. Blue lines are for the non-adapted case and the red lines are for the adapted case. Dashed lines denote the CT safe set boundaries.}% and dotted lines denote the $\phi_\theta=\sigma_{\text{DT}}$ safety margins.}
    \label{fig:phi_traj}
\end{figure}

In (\ref{eq:safety_index}), observe that 
% $k$ is the only parameter that SIS and SIA are finding. A
a larger $k$ value forces the safe control law to be more pre-emptive, e.g., when the quadruped is approaching an obstacle at high speed, a larger $k$ value triggers the $\phi_\theta \ge 0$ condition in (\ref{eq:safe_control_law}) earlier than a lower $k$ value would. 
In Fig.~\ref{fig:phi_traj}, notice that the red solid line ($\phi_{3.5}$ with $k=0.64608$) starts moving away from CT boundary earlier and faster than the blue solid line ($\phi_{0.0}$ with $k=0.61068$). Although $\phi_{3.5}$ also invades the CT boundary, its magnitude is much smaller than $\phi_{0.0}$ and it converges back into the safe region much faster.
This allows the adapted safety index to shift away from unsafe regions earlier and faster, illustrating how SIA adapts the safety index to the varying dynamics.
% which is essential when carrying heavier payloads since its added weight causes momentum. 
% This illustrates how SIA adapts the safety index to the varying dynamics.
% While our work does not yet demonstrate the FI guarantee for (\ref{eq:affine_varying_dynamics}) in real-world scenarios, it showcases SIA's effectiveness in real-time deployment for safe control under varying dynamics. Specifically, SIA ensures that the FTC condition is met promptly, even in challenging conditions with varying payloads on a quadruped. Future work will focus on developing and implementing a safe control law that guarantees FI for discrete-time systems.